\documentclass[sigconf,natbib,nonacm]{acmart}

\usepackage{tabularx}

\newcommand{\ie}{\textit{i.e.,} }
\newcommand{\eg}{\textit{e.g.,} }

\begin{document}

\title{Timestamps as Prompts for Geography-Aware \\ Location Recommendation}

\author{Yan~Luo$^{1,2}$, Haoyi~Duan$^3$, Ye~Liu$^1$, Fu-lai Chung$^1$}
\affiliation{
\institution{$^1$Department of Computing, The Hong Kong Polytechnic University}
\institution{$^2$Media Lab, Massachusetts Institute of Technology}
\institution{$^3$College of Computer Science and Technology, Zhejiang University}}
\affiliation{\{\href{mailto:csyluo@comp.polyu.edu.hk}{csyluo},\href{mailto:cskchung@comp.polyu.edu.hk}{cskchung}\}@comp.polyu.edu.hk, \href{mailto:howie@zju.edu.cn}{howie@zju.edu.cn}, \href{mailto:coco.ye.liu@connect.polyu.hk}{coco.ye.liu@connect.polyu.hk}}

\renewcommand{\shorttitle}{Timestamps as Prompts for Geography-Aware Location Recommendation}
\renewcommand{\shortauthors}{Luo et al.}

\begin{abstract}
Location recommendation plays a vital role in improving users' travel experience. The timestamp of the POI to be predicted is of great significance, since a user will go to different places at different times. However, most existing methods either do not use this kind of temporal information, or just implicitly fuse it with other contextual information. In this paper, we revisit the problem of location recommendation and point out that explicitly modeling temporal information is a great help when the model needs to predict not only the next location but also further locations. In addition, state-of-the-art methods do not make effective use of geographic information and suffer from the hard boundary problem when encoding geographic information by gridding. To this end, a Temporal Prompt-based and Geography-aware (TPG) framework is proposed. The temporal prompt is firstly designed to incorporate temporal information of any further check-in. A shifted window mechanism is then devised to augment geographic data for addressing the hard boundary problem. Via extensive comparisons with existing methods and ablation studies on five real-world datasets, we demonstrate the effectiveness and superiority of the proposed method under various settings. Most importantly, our proposed model has the superior ability of interval prediction. In particular, the model can predict the location that a user wants to go to at a certain time while the most recent check-in behavioral data is masked, or it can predict specific future check-in (not just the next one) at a given timestamp.
\end{abstract}


\begin{teaserfigure}
\vspace{0.1cm}
\centering
\includegraphics[width=0.9\textwidth]{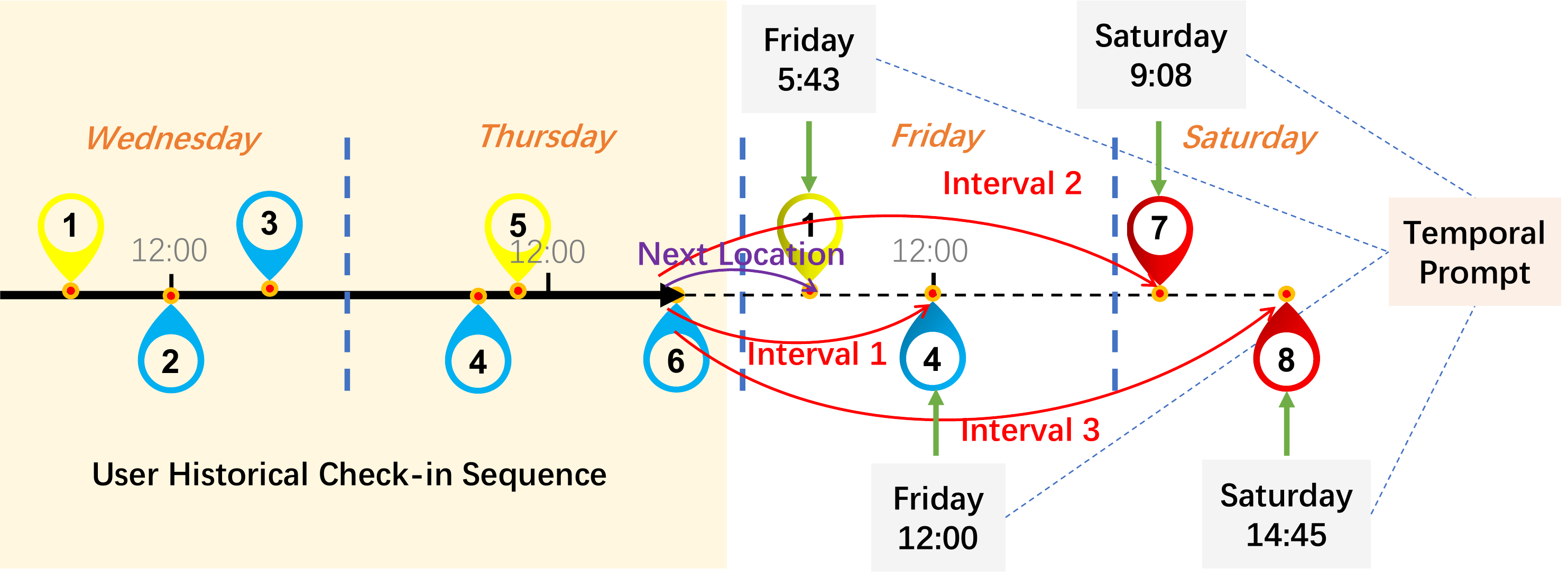}
\caption{An illustration of how TPG performs the next location recommendation (denoted by the purple line) and interval predictions (denoted by red lines) by using temporal prompts. Different colored markers denote different categories of POIs.}
\label{fig:1}
\vspace{0.3cm}
\end{teaserfigure}

\maketitle

\section{Introduction}

Location recommendation is a product of necessity in the era of big data. On the one hand, the prevalence of location-based social networks (LBSNs), like Foursquare and Gowalla, has led to a tremendous amount of user check-in data. On the other hand, while a typical city has thousands of POIs, most users visit very limited POIs in and out of his/her hometown \cite{song2010modelling,luo2022characterizing}. By exploiting user check-in data, location recommendation can provide people with their most interesting locations out of numerous POIs. Location recommendation can not only satisfy a user's travel needs but also has great commercial value for applications such as precision advertising.

A large number of efforts have been devoted to improving the quality of location recommendation. Some studies adopt the most popular solution used by recommender systems, matrix factorization \cite{zheng2010collaborative,lian2014geomf}. However, they do not consider inherent spatial-temporal sequential behaviors and hence achieve suboptimal performance. Traditional statistics-based Markov chain models \cite{baumann2013influence,gambs2012next} have been widely adopted to solve this sequential recommendation problem, but they have limitations, such as only considering the influence from the last check-in activity. With the increase of data volume and the development of deep learning technology, recurrent neural network (RNN) based methods \cite{zhu2017next,feng2018deepmove} are employed to consider long historical information with much stronger representation ability than Markov chain-based models. To tackle the sparsity issue in matrix factorization-based models, some graph-based methods \cite{xie2016learning,wang2022graph} are proposed and can achieve great performances. Recently, the attention mechanism has been proposed and can achieve impressive performance to model long-range temporal and spatial dependencies in human trajectories for location recommendation \cite{lian2020geography,xue2021mobtcast}. As for inputs of location recommender systems, except simple check-in sequences, scholars also try many information sources, such as social relationships and geography information \cite{xue2021mobtcast}.

\begin{figure}
\vspace{0.15cm}
\centering
\includegraphics[width=0.9\linewidth]{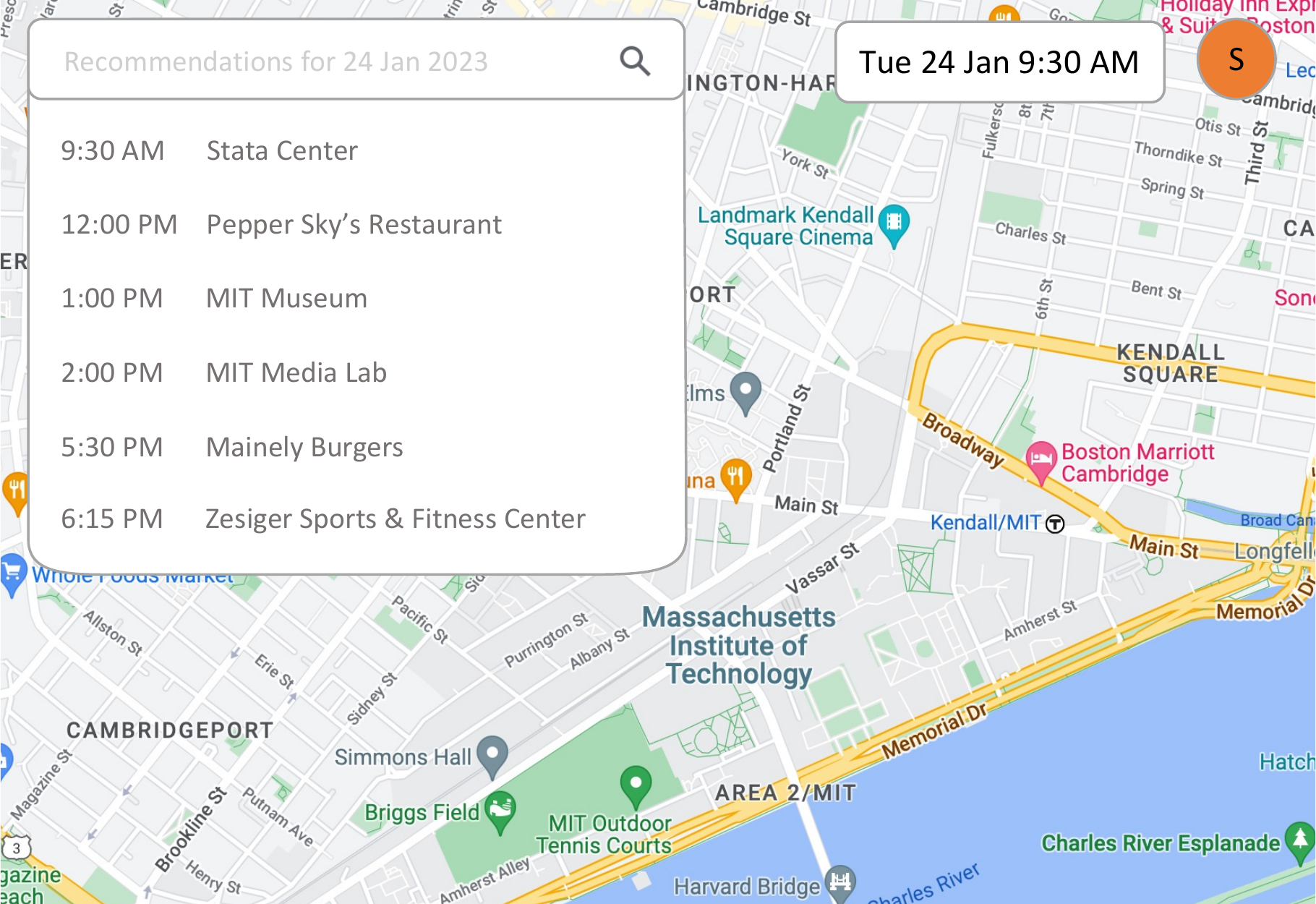}
\caption{One possible use case of TPG is in map services. The left column is a recommendation list generated by TPG for multiple timestamps. The upper right orange circle displays the user information. The bar to the left of the orange circle is the current timestamp.}
\label{fig:2}
\end{figure}

Nevertheless, there are two main challenges not yet addressed among these methods. The first one is that most context-aware state-of-the-art methods implicitly incorporate the temporal information associated with check-ins. They fused time interval information or timestamps of check-ins together with check-in history. This results in inflexible and inaccurate POI prediction outputs when the model needs to predict not only the next location but also further locations. For example, STAN \cite{luo2021stan} builds spatial-temporal matrices for all check-ins within the trajectory slice. CARA \cite{manotumruksa2018contextual} leverages both sequences of check-ins and contextual information associated with the sequences. If we want to predict the 101-st check-in and the 102-nd check-in based on the first 100 historical check-ins, the inputs (\ie the first 100 historical check-ins and their associated contextual information) for both the 101-st and the 102-nd check-in prediction are the same. In other words, these models do not consider different timestamps of the locations to be predicted. However, the fact is that people will tend to visit different locations at different times. For example, a user usually goes to his/her work in the morning and returns home in the evening. Thus, it is important for location recommender systems to have the ability to generate specific prediction(s) based on certain timestamp(s). It is worth noting that methods such as STAN and CARA of course can predict the 102-nd check-in by involving the 101-st predicted check-in into the first 100 historical check-ins. However, in this way this type of models all need to re-train themselves, which will require a large computational overhead.

The second main challenge is that the geographic information is very important in location recommendation, since it is a spatial-temporal problem. State-of-the-art methods do not make effective use of geographic information. MobTCast \cite{xue2021mobtcast} directly feeds the latitude and longitude of POIs into an encoder. However, since check-in data is extremely sparse \cite{adomavicius2005toward}, processing geographic information in this way makes it difficult to capture the physical proximity and dependency between locations. GeoSAN \cite{lian2020geography} further proposes to use hierarchical grids to model the spatial clustering phenomenon in human mobility. However, it suffers from the hard boundary problem, meaning that the POIs near the grid boundary are manually separated.

To tackle these issues, we propose a Temporal Prompt-based and Geography-aware (TPG) framework for location recommendation. 
We make TPG a Transformer-based framework, because Transformer \cite{vaswani2017attention} is originally designed for sequential data with uncertain length, and can differentiate the informativeness of different check-ins and aggregate all check-ins in the trajectory simultaneously for prediction.
Firstly, a geography-aware encoder is designed to capture the geographic correlations among POIs. To avoid the aforementioned hard boundary problem, we propose a shifted window mechanism in the geography-aware encoder. It can bridge the proximity gaps between two adjacent grids and connect adjacent grids by aggregation. Subsequently, the information of user, POI, time, and geography from historical check-in sequences are incorporated by a history encoder, which is designed to learn a comprehensive representation of user travel preference. Afterwards, by using a timestamp as a prompt and regarding it as the query, a temporal prompt-based decoder is utilized to predict the future location(s). In this way, TPG explicitly incorporates the timestamp of the location to be predicted, separating historical check-in sequences and timestamp information. Thus, TPG is very flexible with respect to multiple scenarios. It can not only perform next location recommendations, but also handle interval predictions by using temporal prompts based on those future locations. It should be noted that there are two equivalent scenarios for interval prediction: (a) predicting some further check-ins (\eg the 102-nd, the 103-rd) based on a fixed length of history trajectory slice (\eg the first 100 check-ins), (b) predicting a future location (\eg the 100-th) while the most recent check-in behavioral data being masked (\eg using the first 95 or 96 check-ins). Figure~\ref{fig:1} is a simple example demonstrating how TPG performs the next location recommendation and interval prediction. As shown in Figure~\ref{fig:1}, given the user historical check-in sequence is POI 1-6 from Wednesday to Thursday, the model can know the next four locations the user will visit are POI 1 at 5:43 Friday, POI 4 at 12:00 Friday, POI 7 at 9:08 Saturday, and POI 8 at 14:45 Saturday. Predicting POI 1 at 5:43 Friday is the task of next location recommendation. By making use of temporal prompts, TPG can also predict the location that a user wants to go at a certain time (\ie interval prediction). For example, the model can predict POI 4 at 12:00 Friday (interval 1), POI 7 at 9:08 Saturday (interval 2), and POI 8 at 14:45 Saturday (interval 3), only based on historical check-in sequence POI 1-6. Figure~\ref{fig:2} is one possible use case for a real-world application.

To summarize, the contributions of this paper can be listed as follows:
\begin{itemize}
\item We argue that the explicitly modeling timestamp of the location to be predicted is essential in real-world applications. A novel and effective Transformer-based framework named TPG is proposed. Temporal information is regarded as a prompt for our recommendation system.
\item To effectively utilize geographic information, we propose a geography-aware encoder with a shifted window mechanism devised to avoid the hard boundary problem when treating longitude and latitude of POIs with grids.
\item Experimental results on five real-world datasets, namely, Gowalla, Brightkite, Foursquare-NYC, Foursquare-TKY, and Foursquare-SIN, show that our model outperforms the state-of-the-art counterparts under different settings. We also demonstrate that TPG's interval prediction perform much better than baselines.
\end{itemize}

\section{Related Work}

In this section, we firstly review recent works on sequential recommendation and next location recommendation. Then, we further describe the recent advance in prompt-based learning in the field of natural language processing (NLP). 

\subsection{Sequential Recommendation}

Sequential recommendation models, initially based on Markov chain \cite{baumann2013influence,gambs2012next}, were able to predict probability of the next behavior via a transition matrix. Challenged with the flaw of Markov-based model that only the transition probability between two consecutive visits is mainly considered, deep learning-based models, among which RNN \cite{zhang2014sequential} models are representative and have been quickly developed as strong baselines, go on stage. Meanwhile, the contributions of other deep learning methods, such as metric embedding algorithms \cite{feng2020hme}, convolutional neural networks \cite{tang2018personalized}, reinforcement learning algorithms \cite{massimo2018harnessing}, and graph neural networks \cite{wu2019session,yu2020tagnn} can be witnessed. Recently, due to full parallelism and the capacity of capturing long-range dependence, the self-attention networks \cite{vaswani2017attention} are proposed for sequential recommendation and have achieved distinguished performances. For example, GeoSAN \cite{lian2020geography} employs self-attention layers and achieves state-of-the-art performance.

However, most of the methods mentioned predict the next location only through the fusion of historical check-in sequences and associated contextual information, ignoring that the timestamp of the location to be predicted is also of great significance. The proposed TPG framework explicitly models temporal information via a novel temporal prompt.

\subsection{Next Location Recommendation}

Next location recommendation, \ie POI recommendation, which draws considerable attention recent years due to great business value, can be viewed as a special sub-task of sequential recommendation with spatial information \cite{luo2021stan}. Regarding the use of spatio-temporal information in next location recommendation, many previous works only use spatio-temporal intervals between two successive visits in a recurrent layer. For example, DeepMove \cite{feng2018deepmove} combines an attention layer for learning long-term sequential regularity. LSTPM \cite{sun2020go} proposes a geo-dilated RNN that aggregates locations visited recently, but only for shot-term preference. Inspired by sequential item recommendation \cite{kang2018self}, GeoSAN \cite{lian2020geography} uses self-attention model in next location recommendation within the trajectory. STAN \cite{luo2021stan} adopts a spatial-temporal attention network that aggregates all relevant check-ins in trajectories. GETNext \cite{yang2022getnext} proposes a novel graph enhanced Transformer model by exploiting the extensive collaborative signals.

However, these models suffer from limitations in geographic and temporal information modeling. In TPG, we take the challenge of the hard boundary problem in grid mapping, and propose the shifted window mechanism. We also discard the implicit way to fuse the temporal information in TPG. Temporal prompt is proposed for explicitly modeling timestamp of locations to be predicted.

\begin{figure*}
\centering
\includegraphics[width=0.85\linewidth]{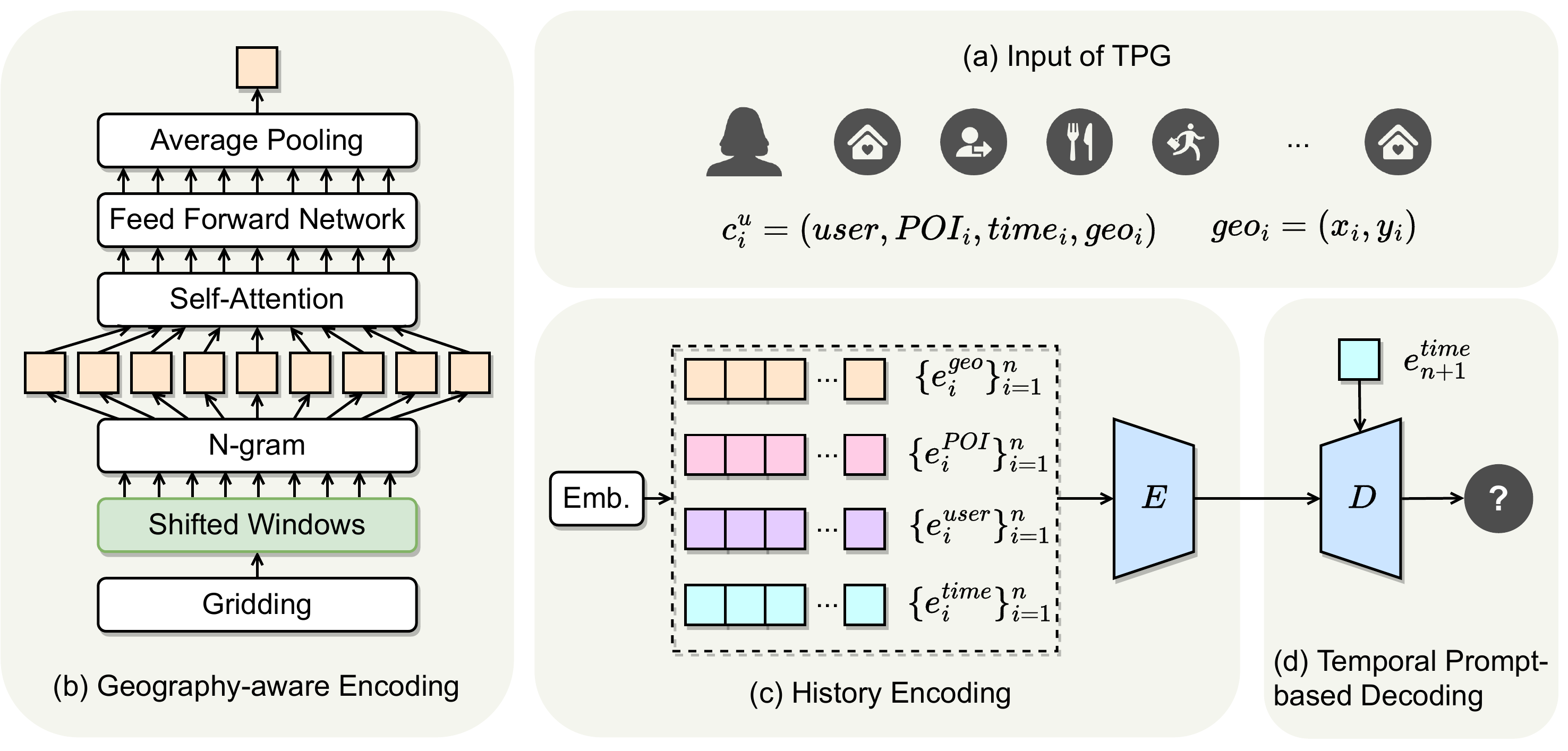}
\caption{The overall architecture of the proposed TPG. Detailed explanations of notations are described in Section~\ref{sec:3}.}
\label{fig:3}
\end{figure*}

\subsection{Prompt-based Learning}

In recent years, we have witnessed the success of pre-trained language models, such as GPT series especially GPT3 \cite{brown2020language}, which has triggered prompt-based learning paradigm, namely “pre-train, prompt, and predict” \cite{liu2023pre}, to boom on NLP tasks. At this point, models are no longer designed to adapt to the tasks. Instead, various tasks are modified by templates to get closer to the space of pre-trained model, which is trained with huge language data from the web. During the whole developing process, many prompt design methods have proliferated. One type of the methods \cite{gao2020making} is proper discrete prompts, concentrating on the design of the templates. Meanwhile, another line of work \cite{liu2021p} exploited continuous vector embeddings as prompts to improve the performance on NLP tasks. Furthermore, as yet another type, instruction-based prompts contain detail task descriptions and adhere more to the natural language format. 

As a matter of fact, recommendation has been co-evolved with NLP techniques for a long time. Inspired by the success of these approaches on NLP tasks, researchers recently transfer the main idea of prompt-based learning to recommendation problems, achieving very promising results. P5 \cite{geng2022recommendation} creates a collection of personalized prompts and then trains a sequence-to-sequence model on a variety of recommendation related tasks verbalized according to the constructed personalized prompts. 

To our best knowledge, our work is the first one to use timestamps as prompts in the self-attention based sequential recommendation framework for location recommendation, which explicitly models temporal information.

\section{New Framework}\label{sec:3}

In this section, more details about the proposed TPG framework are elaborated. We first give the problem statement and provide an overview of the framework. Then, we elaborate on the three main modules of TPG, \ie geography-aware encoder, history encoder, and temporal prompt-based decoder.

\subsection{Overview}
Each check-in $c_{i}^u = (u, t_i, p_i)$ is a user, POI, time tuple, which denotes a behavior that a user $u$ visits POI $p_i$ at time $t_i$. Each POI $p_i$ has its own geographic coordinates $(x_i, y_i)$. Each user $u$ has a sequence of historical check-ins $C_{1 \rightarrow n}^u = \{c_{i}^u\}_{i=1}^n$. Given the historical check-in sequences of users, the goal of next location recommendation is to predict the next POI $\rho_{t_{n+1}}$ that a certain user $u$ will visit at a certain time $t_{n+1}$.

The overall architecture of our TPG framework is described in Figure~\ref{fig:3}. Based on the Transformer's encoder-decoder structure, TPG can be divided into three parts, \ie geography-aware encoder, history encoder, and temporal prompt-based decoder. For each check-in, the geographic coordinate of POI can be fed into the geography-aware encoder to get geographical representation $e_i^{geo}$. The historical check-in sequences including POI, user, and time information are then fed into the multi-modal embedding module to generate hidden representations $\{e_i^{POI}\}_{i=1}^n$, $\{e_i^{user}\}_{i=1}^n$, and $\{e_i^{time}\}_{i=1}^n$. Together with $\{e_i^{geo}\}_{i=1}^n$ from the geography-aware encoder, these representations are processed by a history encoder to generate user travel preference representation. Using temporal information of $t_{n+1}$ as prompt, the temporal prompt query and user travel preference memory are then forwarded to the decoder, which is capable of generating more accurate predictions for the next locations.

\subsection{Geography-aware Encoder}

\begin{figure}
\vspace{0.1cm}
\centering
\includegraphics[width=0.7\linewidth]{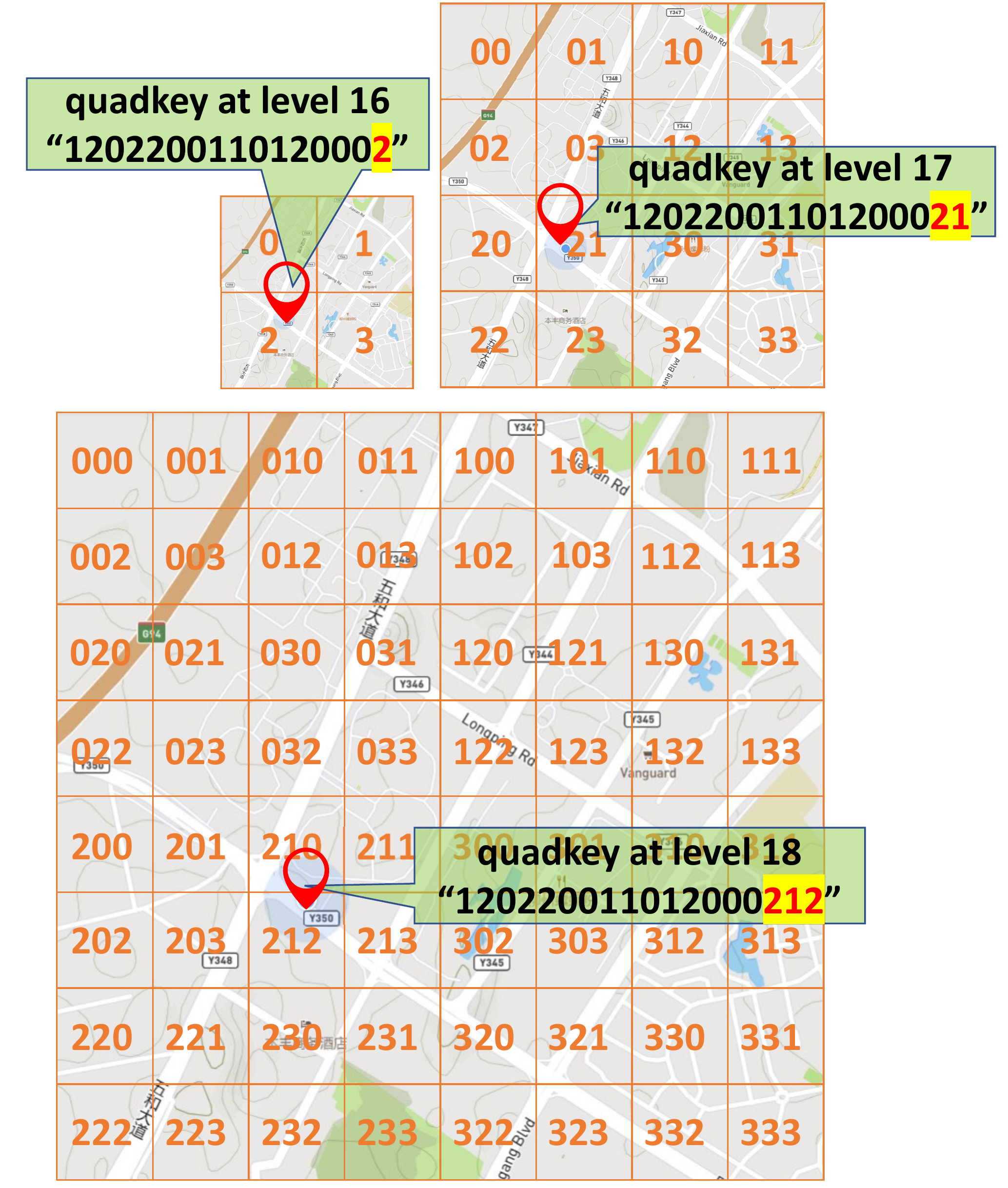}
\caption{An illustration of hierarchical gridding based on the tile map system. The given example is about mapping a location into grids at level 16-18, whose quadkeys are annotated.}
\label{fig:4}
\end{figure}

Sparsity issue is a key challenge in recommendation problem. In particular, as the check-in data gives implicit feedback of visiting behavior \cite{lian2014geomf}, the data sparsity problem is even worse for POIs compared other item recommendation such as movies or goods, of which users usually only express their opinion with ratings. Thus, when it comes to encoding geographic information of POIs, directly feeding the coordinates of POIs into the learning model makes it difficult for the model to capture geographic correlations. GeoSAN \cite{lian2020geography} embeds the exact position of locations by mapping latitude and longitude into hierarchical grids using tile map system\footnote{\url{https://www.maptiler.com/google-maps-coordinates-tile-bounds-projection}} as exemplified in Figure~\ref{fig:4}. The tile map system is a hierarchical multi-resolution pyramid model. The world map is obtained by projecting the entire world into a flat plane by Mercator\footnote{\url{https://www.britannica.com/science/Mercator-projection}}. The scale of the plane starts with $512 \times 512$ pixels. It grows by a factor of 2 with the increase of levels. For better retrieval and display, the plane is further divided into grids of $256 \times 256$ pixels each. From the low level bottom to the high level top of the tile pyramid, the resolution becomes lower and lower, but the geographic range is unchanged via sub-gridding one grid into four grids of the same size. Since the partition of grids is like quadtree, each grid can be identified with a unique quadtree key (quadkey for short). Quadkeys consist of the characters from the set \{``0'', ``1'', ``2'', ``3''\}. The length of it equals the level of grid.

\begin{figure}
\centering
\includegraphics[width=0.35\linewidth]{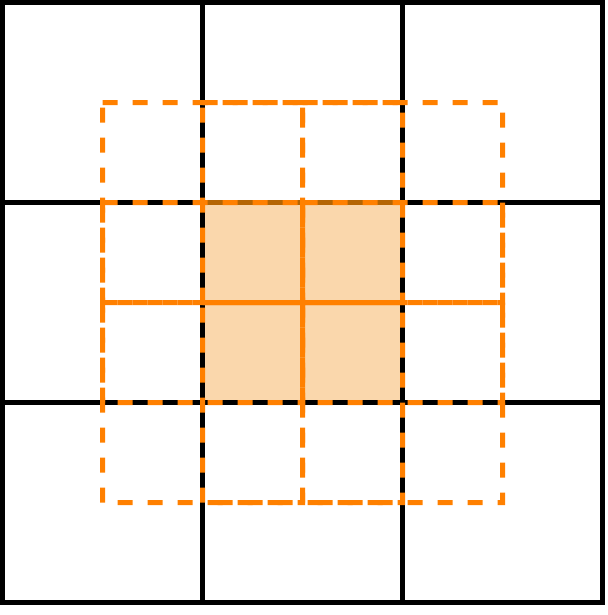}
\caption{An illustration of shifted windows for the grid marked by orange. Here, we take the rolling step as half of the length of the grid size.}
\label{fig:5}
\end{figure}

It indeed can alleviate the sparsity problem to some extent. However, for grids at the same level, the boundary of grids may damage the physical spatial proximity of two POIs around the boundary, which is a violation of Tobler's First Law of Geography \cite{tobler1970computer}. In other words, lacking connections across adjacent grids limits modeling power of geography-aware encoder. To introduce cross-grid connections, we propose a shifted window mechanism in our geography-aware encoder. As illustrated in Figure~\ref{fig:5}, for each grid, we move the shifted window along the X and Y direction (and both) by a certain step, which is part of the length of the grid size. In this way, we will get nine grids for each grid, \ie itself and eight augmented neighbor grids.

Now the remaining task of geography-aware encoder is transforming quadkeys of these nine grids into a continuous latent embedding with rich information. For each quadkey, it is actually a character sequence. Each character in the sequence denotes the index of the grid partition at a certain level. If we want to make effective use of hierarchical spatial information of grids, an intuitive and straightforward approach is to conduct self-attention between these characters. However, since the cardinality of the character set is very small (\ie only 4), treating a quadkey at character-level cannot achieve the goal of fully encoding the geographic correlations between POIs. Therefore, we consider dividing the character sequence by n-gram, and converting it into a sequence at n-gram-level. In this way, the vocabulary size of the sequence increases from 4 to 4n. For example, if a quadkey is ``013201233'', the result of using four-gram is 0132-1320-3201-2012-0123-1233. We then use a stacked self-attention network and a point-wise feed forward network for capturing dependencies among these n-grams. After that, for each grid among these nine grids, average pooling can be utilized to aggregate the sequence of n-gram representations. We then obtain geographic embedding $e_i^{geo}$ for the given location $POI_i$ by average pooling on aggregated representations of these nine grids. The shifted window mechanism is helpful to reduce possible bias caused by the arbitrary partition of grids via including information from augmented neighbor grids.

\subsection{History Encoder}

Each check-in is a tuple consisting of user, time and POI information. To tackle with such discrete and heterogeneous data, we need a multi-modal embedding module to transfer check-in data into interpretable information for TPG. Specifically, for encoding time information, timestamps are firstly discretized into $24 \times 7 = 168$ types learnable vectors. For encoding user and POI information, the type number of learnable vectors equals to the unique number of users and POIs in datasets. All these vectors are then linearly projected into $d$-dimensional embeddings $e_i^{time} \in \mathbb{R}^d$, $e_i^{user} \in \mathbb{R}^d$, and $e_i^{POI} \in \mathbb{R}^d$. In this way, for the user $u$, the historical check-in sequence $\{c_{i}^u\}_{i=1}^n$ can be further denoted as $(\{e_i^{POI}\}_{i=1}^n, \{e_i^{user}\}_{i=1}^n, \{e_i^{time}\}_{i=1}^n, \{e_i^{geo}\}_{i=1}^n)$. Note that since check-in data requires a certain order of precedence, learnable positional embedding is also added into inputs for history encoder.

Compared with previous RNN-based methods, Transformer architecture \cite{vaswani2017attention} can not only avoid recurrence, allowing parallel computing to reduce training time, but also migrate performance degradation problem with regard to long-term dependencies in RNNs. To better capture long range spatial-temporal dependencies in users' historical check-in sequences, we stack Transformer encoder layers \cite{vaswani2017attention} for constructing the history encoder. Each Transformer encoder layer involves a multi-head self-attention module and a point-wise feed-forward network. We also keep the residual connection and layer normalization employed in Transformer encoder layers. Dividing the attention mechanism into multiple heads to form multiple sub-spaces allows the model to focus on different aspects of information. For each attention head, self-attention result for a check-in $c_i$ can be computed as
\begin{gather}\label{eq:1}
ATTENTION(e^c_i) = w_z \sum_{j=1}^{N_v} \frac{\exp(w_q e^c_i \times w_k e^c_j)}{\sum_{m=1}^{N_v} \exp(w_q e^c_i \times w_k e^c_m)} w_v e^c_j + e^c_i
\end{gather}
where $e^c_i$ is the input check-in embedding for $c_i$, $e^c_j$ is the embedding for contextual check-in in the sequence, and $w_{\{q, k, v, z\}}$ denotes linear transform weights for the query, key, value, and output matrices. The self-attention mechanism aggregates the global context information into each check-in features. After multi-head self-attention results are obtained by concatenating every self-attention result, the encoder is able to jointly attend to information from different representation sub-spaces at different positions. Then, the feed-forward network contains two linear transformations with a ReLU activation in between, which can be denoted as
\begin{gather}\label{eq:2}
FFN(e^c_{i'}) = max(0, e^c_{i'}W_1+b_1)W_2+b_2
\end{gather}
where $e^c_{i'}$ is the input embedding after multi-head self-attention, $W_1$ and $W_2$ denote linear transform weights, and $b_1$ and $b_2$ are linear transform offsets.

\subsection{Temporal Prompt-based Decoder}

The normal way for existing methods to generate the predictions of next location is based on historical check-ins and the associated contextual information. The model will not explicitly indicate the output with respect to a special temporal information. However, in real world applications, it is important to consider the exact time for next location prediction. Location recommendation is usually employed in map services such as Google Map and location based services such as Foursquare. We can regard the time of user opening the app or clicking the query box in the app as the timestamp of next location. Human mobility has the periodicity \cite{yuan2014graph}. People will revisit POIs of the same category around the same hour of different days. While at different hour of a day, people tend to visit different types of POIs. For example, it is the simple fact that if a user clicking the query box in the map app at noon, he/she is probably looking for a restaurant. While in the morning, there is a big chance that he/she wants to search for the route to his/her company. Therefore, if the model does not know the timestamp of next location, it cannot produce results of next locations with high confidence. Thus, an intuitive idea is that we can directly tell the model about the timestamp of next location. A simple method for it is to incorporate this timestamp into inputs. However, since a check-in is a tuple of user, time, and POI, directly adding this single timestamp into inputs is not appropriate. To this end, we propose a temporal prompt-based decoder, using timestamps as prompts and queries for the decoder. There are several advantages of utilizing temporal prompts: (1) It separates historical check-in sequences and temporal information of locations to be predicted, making the model more flexible for generating any future check-in. (2) By explicitly modeling temporal information, the prediction is greatly correlated to the given timestamp. As mentioned above, people's travel choices are strongly related to timestamps, so this design will most likely improve model performances.

In greater detail, the preference decoder takes queries (\ie time representation $e_{n+1}^{time}$) and encoder memory (\ie user travel preference representation $e^{C}$) as inputs. We construct the temporal prompt-based decoder by stacking Transformer decoder layers\cite{vaswani2017attention}, each of which consists of a multi-head self-attention sub-layer, an encoder-decoder attention sub-layer, and a feed-forward network sub-layer. User travel preferences and timestamp information are deeply fused in each encoder-decoder attention sub-layer. Each attention head of the encoder-decoder attention sub-layer can be represented by
\begin{gather}\label{eq:3}
ATT(e^C,e_{n+1}^{time}) = w_z \sum_{j=1}^{N_v} \frac{\exp(w_q e^C \times w_k e_{n+1}^{time})}{\sum_{m=1}^{N_v} \exp(w_q e^C \times w_k e_{n+1,m}^{time})} w_v e_{n+1}^{time} + e^C
\end{gather}
where notations are consistent with Eq.~\ref{eq:1}.

After adding up and feeding the results into the feed-forward networks for further projection, the output embedding decodes the fused check-in features and has the same length as the query embedding. The output here is actually the embedding of predicted next location.

As for the training scheme, we adopt the negative log likelihood with sampled Softmax as the recommendation loss for each user $u$. The recommendation loss can be depicted as:
\begin{equation}
\mathcal{L}_{rec}(\tilde{y}) = -\log\frac{\exp(\tilde{y}y^+)}{\exp(\tilde{y}y^+) + \displaystyle{\sum_{y^- \in \mathcal{Y}^-}}\exp(\tilde{y}y^-)}
\end{equation}
where $\tilde{y}$, $y^+$, and $y^-$ indicate the inferred location embedding, the ground truth of location which user $u$ visits at time $t_{n+1}$, and the randomly sampled negative data which user does not visit at $t_{n+1}$, respectively.

\section{Experiments and Evaluations}

In this section, we report the extensive experiments conducted to evaluate the performance and show the superior performance of our method. The results also demonstrate the effectiveness and utility of TPG. Further experiments also validate the rationality of each component of TPG.

\subsection{Experimental Settings}

\subsubsection{Datasets}

We use five publicly available real-world Location-Based Social Network datasets to evaluate our method: Gowalla\footnote{\url{https://snap.stanford.edu/data/loc-gowalla.html}}, Brightkite\footnote{\url{http://snap.stanford.edu/data/loc-brightkite.html}}, NYC, TKY\footnote{\url{https://drive.google.com/file/d/0BwrgZ-IdrTotZ0U0ZER2ejI3VVk/view?usp=sharing&resourcekey=0-rlHp_JcRyFAxN7v5OAGldw}}, and SIN\footnote{\url{https://www.ntu.edu.sg/home/gaocong/data/poidata.zip}}. Gowalla and Brightkite contain worldwide data while the NYC, TKY, and SIN are extracted from Foursquare global dataset, which only focuses on a single city/region. We adopted LibCity's \cite{wang2021libcity} pre-processing pipeline. Table~\ref{tab:1} gives a rough sketch of the statistics of the five datasets.

\begin{table}
\setlength\tabcolsep{3.5pt}
\centering
\caption{Dataset statistics.}
\label{tab:1}
\begin{tabularx}{\linewidth}{p{1.6cm}<{\raggedright}ccccc}
\toprule
& Gowalla & Brightkite & NYC & TKY & SIN \\
\midrule
\#users & 31,708 & 5,247 & 1,010 & 6,771 & 367 \\
\#locations & 131,329 & 48,181 & 5,135 & 14,590 & 3,104 \\
\#check-ins & 2,963,373 & 1,699,579 & 140,229 & 871,200 & 136,847 \\
\bottomrule
\end{tabularx}
\end{table}

\subsubsection{Evaluation Metrics}

We adopt two widely-used metrics of ranking evaluation: Recall and NDCG, to evaluate recommendation performance. Recall@k counts the rate of true positive samples in all positive samples, which in our case means the rate of the label in the top-k probability samples, NDCG rewards method that ranks positive items in the first few positions of the top-k ranking list. We report $k$=5 and $k$=10 in our experiments.

\begin{table*}
\setlength{\tabcolsep}{0pt}
\centering
\caption{Comparison with eight baselines. The best performances are boldfaced. The second best performances are underlined.}
\label{tab:2}
\begin{tabularx}{\linewidth}{p{1.45cm}<{\raggedright}p{0.815cm}<{\centering}p{0.815cm}<{\centering}p{0.815cm}<{\centering}p{0.815cm}<{\centering}p{0.815cm}<{\centering}p{0.815cm}<{\centering}p{0.815cm}<{\centering}p{0.815cm}<{\centering}p{0.815cm}<{\centering}p{0.815cm}<{\centering}p{0.815cm}<{\centering}p{0.815cm}<{\centering}p{0.815cm}<{\centering}p{0.815cm}<{\centering}p{0.815cm}<{\centering}p{0.815cm}<{\centering}p{0.815cm}<{\centering}p{0.815cm}<{\centering}p{0.815cm}<{\centering}p{0.815cm}<{\centering}}
\toprule
& \multicolumn{4}{c}{Gowalla} & \multicolumn{4}{c}{Brightkite} & \multicolumn{4}{c}{NYC} & \multicolumn{4}{c}{TKY} & \multicolumn{4}{c}{SIN} \\ 
\cmidrule(r){2-5} \cmidrule(r){6-9} \cmidrule(r){10-13} \cmidrule(r){14-17} \cmidrule{18-21}
\multicolumn{1}{c}{} & R@5 & N@5 & R@10 & N@10 & R@5 & N@5 & R@10 & N@10 & R@5 & N@5 & R@10 & N@10 & R@5 & N@5 & R@10 & N@10 & R@5 & N@5 & R@10 & N@10 \\
\midrule
HSTLSTM & 44.6\% & 30.9\% & 54.4\% & 35.1\% & 46.8\% & 37.3\% & 50.4\% & 41.3\% & 25.7\% & 19.7\% & 31.1\% & 21.4\% & 29.1\% & 22.1\% & 35.7\% & 24.3\% & 14.9\% & 11.2\% & 21.0\% & 13.4\% \\
DeepMove & 49.9\% & 35.9\% & 59.1\% & 39.0\% & 50.3\% & 39.1\% & 58.7\% & 43.2\% & 29.8\% & 21.6\% & 36.5\% & 23.8\% & 33.3\% & 25.1\% & 39.8\% & 27.2\% & 16.8\% & 11.8\% & 24.7\% & 13.4\% \\
LSTPM & 42.2\% & 30.4\% & 53.2\% & 33.0\% & 43.8\% & 34.9\% & 52.4\% & 38.5\% & 22.8\% & 16.1\% & 30.7\% & 18.6\% & 37.2\% & 28.2\% & 45.4\% & 30.9\% & 13.1\% & 9.2\% & 20.9\% & 12.0\% \\
TMCA & 44.3\% & 32.5\% & 55.4\% & 35.5\% & 45.5\% & 36.4\% & 55.9\% & 41.6\% & 24.6\% & 18.3\% & 33.1\% & 20.2\% & 39.9\% & 27.9\% & 41.7\% & 29.5\% & 15.5\% & 11.3\% & 22.3\% & 14.0\% \\
CARA & 50.2\% & 36.6\% & 60.0\% & 40.8\% & 51.6\% & 40.2\% & 54.7\% & 41.2\% & 28.0\% & 20.2\% & 37.5\% & 24.0\% & 31.8\% & 24.3\% & 37.2\% & 28.0\% & 15.04\% & 11.8\% & 20.8\% & 14.0\% \\
MobTCast & 54.3\% & 37.9\% & 65.5\% & 46.6\% & 52.5\% & 43.6\% & 59.5\% & 46.4\% & 31.3\% & 21.3\% & 41.3\% & \underline{28.1\%} & 59.7\% & 48.4\% & 65.4\% & 52.6\% & 17.0\% & \underline{12.4\%} & 24.6\% & 15.1\% \\
STAN & \underline{58.7\%} & \underline{41.6\%} & \underline{70.3\%} & \underline{46.6\%} & \underline{57.2\%} & \underline{45.4\%} & \underline{69.8\%} & \underline{47.3\%} & \underline{32.1\%} & \underline{22.3\%} & \underline{45.9\%} & 27.3\% & \underline{61.2\%} & \underline{50.1\%} & \underline{71.3\%} & \underline{54.4\%} & \underline{18.0\%} & 11.3\% & 29.5\% & 15.7\% \\
GeoSAN & 56.2\% & 41.4\% & 69.9\% & 45.9\% & 55.8\% & 42.3\% & 67.2\% & 46.0\% & 30.2\% & 20.8\% & 44.5\% & 25.3\% & 58.8\% & 48.4\% & 69.1\% & 51.8\% & 17.4\% & 11.8\% & \underline{30.0\%} & \underline{15.9\%} \\
\midrule
TPG & \textbf{63.2\%} & \textbf{45.5\%} & \textbf{74.6\%} & \textbf{50.1\%} & \textbf{62.4\%} & \textbf{47.9\%} & \textbf{74.4\%} & \textbf{51.8\%} & \textbf{37.3\%} & \textbf{26.8\%} & \textbf{52.8\%} & \textbf{31.8\%} & \textbf{65.4\%} & \textbf{53.1\%} & \textbf{76.5\%} & \textbf{56.7\%} & \textbf{19.4\%} & \textbf{14.3\%} & \textbf{34.3\%} & \textbf{19.1\%} \\
Improv. & 7.7\% & 9.4\% & 6.1\% & 7.5\% & 9.1\% & 5.5\% & 6.6\% & 9.5\% & 16.2\% & 20.2\% & 15.0\% & 13.2\% & 6.9\% & 6.0\% & 7.3\% & 4.2\% & 7.8\% & 15.3\% & 14.3\% & 20.1\% \\
\bottomrule
\end{tabularx}
\end{table*}

\subsubsection{Baselines}

To show the effectiveness of our proposed methods, we compare our proposed TPG with several following baselines:

\begin{itemize}
\item HSTLSTM \cite{kong2018hst}: a LSTM based method which introduces spatio-temporal transfer factors and uses an encoder-decoder structure for prediction.
\item DeepMove \cite{feng2018deepmove}: an attentional recurrent network which capture the complicated sequential transitions and the multi-level periodicity.
\item LSTPM \cite{sun2020go}: a long- and short-term preference modeling framework which consists of a nonlocal network for long-term preference modeling and a geo-dilated RNN for short-term preference learning.
\item CARA \cite{manotumruksa2018contextual}: a novel contextual attention recurrent architecture that leverages both sequences of feedback and contextual information associated with the sequences to capture the users' dynamic preferences.
\item TMCA \cite{li2018next}: a novel temporal and multi-level context attention LSTM-based encoder-decoder framework which is able to adaptively select relevant check-in activities and contextual factors for next POI preference prediction
\item GeoSAN \cite{lian2020geography}: a geography-aware sequential recommender based on the self-attention network that uses hierarchical gridding of GPS locations for spatial discretization and uses self-attention layers.
\item STAN \cite{luo2021stan}: a spatial-temporal attention network that explicitly aggregates all relevant check-ins in trajectories, not only just successive ones.
\item MobTCast \cite{xue2021mobtcast}: a Transformer-based context-aware network combined with a location prediction branch as an auxiliary task. It which captures temporal, semantic, social and geographical contexts. 
\end{itemize}

\subsubsection{Implementation Details}

For the check-in sequence of each user, we take the last check-in record on a previously unvisited location as ground truth in evaluation, and check-in sequence before that for training. The maximum sequence length is set to 100. 

Different from GeoSAN which directly uses ground truth for negative sampling in both train and evaluation setting and may cause label leakage, we consider a practical scenario where each user's next physical position is unknown, and the negative samples have to be drawn from the vicinity of the immediately preceding check-in location. To be more specific, 100 of the 2000 nearest locations from user's current GPS coordinates are chosen randomly as negative samples. Recall and NDCG can then be computed based on the ranking of these 101 locations.

We run all the experiment on NVIDIA V100 GPUs. For our TPG model, we set the dimension of location and region embeddings to 50 respectively, and time embedding to 100. The step size of the shifted windows is set to a quarter of the side length of the grid. We train our model using the Adam optimizer with a learning rate of 0.001 and set the dropout ratio to 0.5. The number of training epochs is set to 50 for all four datasets. For baselines except STAN, we follow their implementation and best settings which they claim in their papers. STAN builds matrices for all historical check-ins of each user, which results in extremely time consuming and memory consuming. Running the original version of STAN caused an out-of-memory (OOM) error on our server with 768GB memory. To test the performance of STAN, we choose to select a part of users to train model at a time and test performance on these users. For NYC and SIN datasets, we use all users. For other datasets, we select the first 2000 users to test model performance due to the large number of users on these datasets.

\begin{table*}
\setlength{\tabcolsep}{0pt}
\centering
\caption{Interval Prediction Performances. We boldface the interval prediction results higher than TPG.}
\label{tab:3}
\begin{tabularx}{\linewidth}{p{1.5cm}<{\raggedright}p{0.815cm}<{\centering}p{0.815cm}<{\centering}p{0.815cm}<{\centering}p{0.815cm}<{\centering}p{0.815cm}<{\centering}p{0.815cm}<{\centering}p{0.815cm}<{\centering}p{0.815cm}<{\centering}p{0.815cm}<{\centering}p{0.815cm}<{\centering}p{0.815cm}<{\centering}p{0.815cm}<{\centering}p{0.815cm}<{\centering}p{0.815cm}<{\centering}p{0.815cm}<{\centering}p{0.815cm}<{\centering}p{0.815cm}<{\centering}p{0.815cm}<{\centering}p{0.815cm}<{\centering}p{0.815cm}<{\centering}}
\toprule
& \multicolumn{4}{c}{Gowalla} & \multicolumn{4}{c}{Brightkite} & \multicolumn{4}{c}{NYC} & \multicolumn{4}{c}{TKY} & \multicolumn{4}{c}{SIN} \\ 
\cmidrule(r){2-5} \cmidrule(r){6-9} \cmidrule(r){10-13} \cmidrule(r){14-17} \cmidrule(r){18-21}
\multicolumn{1}{c}{} & R@5 & N@5 & R@10 & N@10 & R@5 & N@5 & R@10 & N@10 & R@5 & N@5 & R@10 & N@10 & R@5 & N@5 & R@10 & N@10 & R@5 & N@5 & R@10 & N@10 \\
\midrule
STAN & 58.7\% & 41.6\% & 70.3\% & 46.6\% & 57.2\% & 45.4\% & 69.8\% & 47.3\% & 32.1\% & 22.3\% & 45.9\% & 27.3\% & 61.2\% & 50.1\% & 71.3\% & 54.4\% & 18.0\% & 11.3\% & 29.5\% & 15.7\% \\
\midrule
int. 1 & 54.2\% & 36.5\% & 66.5\% & 42.2\% & 52.1\% & 40.6\% & 66.2\% & 43.3\% & 30.0\% & 20.3\% & 42.5\% & 24.5\% & 57.4\% & 46.9\% & 67.7\% & 52.1\% & 16.8\% & 10.3\% & 27.5\% & 14.4.\% \\
int. 2 & 48.2\% & 31.3\% & 62.2\% & 37.0\% & 50.4\% & 40.9\% & 65.2\% & 42.0\% & 24.6\% & 14.5\% & 36.5\% & 18.3\% & 52.5\% & 43.8\% & 64.4\% & 49.4\% & 13.5\% & 8.3\% & 25.5\% & 13.4\% \\
int. 3 & 42.4\% & 28.1\% & 58.4\% & 33.3\% & 47.0\% & 37.5\% & 63.3\% & 40.3\% & 20.3\% & 10.9\% & 32.1\% & 15.6\% & 47.9\% & 40.2\% & 62.5\% & 45.9\% & 12.1\% & 8.5\% & 23.6\% & 12.1\% \\
\midrule
GeoSAN & 56.2\% & 41.4\% & 69.9\% & 45.9\% & 55.8\% & 42.3\% & 67.2\% & 46.0\% & 30.2\% & 20.8\% & 44.5\% & 25.3\% & 58.8\% & 48.4\% & 69.1\% & 51.8\% & 17.4\% & 11.8\% & 30.0\% & 15.9\% \\
\midrule
int. 1 & 54.3\% & 39.5\% & 67.8\% & 43.9\% & 52.1\% & 40.6\% & 66.2\% & 43.3\% & 29.0\% & 18.6\% & 43.4\% & 23.2\% & 56.3\% & 46.3\% & 67.7\% & 49.3\% & 16.4\% & \textbf{12.0\%} & 29.5\% & 14.4\% \\
int. 2 & 53.2\% & 39.6\% & 65.3\% & 41.2\% & 50.4\% & 40.9\% & 65.2\% & 42.0\% & 28.7\% & 18.3\% & 42.6\% & 22.9\% & 54.3\% & 44.8\% & 65.4\% & 47.6\% & 15.2\% & 10.4\% & 28.6\% & 13.4\% \\
int. 3 & 50.1\% & 35.4\% & 62.1\% & 38.5\% & 47.0\% & 37.5\% & 63.3\% & 40.3\% & 25.7\% & 16.9\% & 39.6\% & 21.0\% & 55.3\% & 43.2\% & 65.2\% & 46.2\% & 14.2\% & 9.4\% & 26.4\% & 13.8\% \\
\midrule
TPG & 63.2\% & 45.5\% & 74.6\% & 50.1\% & {62.4\%} & {47.9\%} & {74.4\%} & {51.8\%} & {37.3\%} & {26.8\%} & {52.8\%} & {31.8\%} & {65.4\%} & {53.1\%} & {76.5\%} & {56.7\%} & {19.4\%} & {14.3\%} & {34.3\%} & {19.1\%} \\
\midrule
int. 1 & \textbf{63.7\%} & 44.3\% & 73.7\% & 49.1\% & 60.4\% & 46.2\% & 73.0\% & 50.3\% & \textbf{38.0\%} & 26.7\% & \textbf{53.6\%} & \textbf{32.1\%} & 64.5\% & 52.9\% & 75.6\% & 56.4\% & \textbf{19.9\%} & 13.3\% & 33.1\% & 18.1\% \\
int. 2 & 60.5\% & \textbf{45.9\%} & 72.8\% & 48.6\% & 59.9\% & 45.6\% & 72.5\% & 49.9\% & \textbf{37.5\%} & 26.5\% & \textbf{53.2\%} & 30.1\% & 64.1\% & 52.2\% & 75.1\% & 55.8\% & 19.1\% & 12.6\% & 32.2\% & 18.7\% \\
int. 3 & 59.7\% & \textbf{46.5\%} & 72.0\% & 47.1\% & 59.8\% & 45.3\% & 72.3\% & 49.8\% & 37.0\% & 25.3\% & 52.0\% & 29.7\% & 64.9\% & \textbf{53.1\%} & 75.1\% & \textbf{56.7\%} & \textbf{19.6\%} & \textbf{14.3\%} & 31.3\% & 17.6\% \\
\bottomrule
\end{tabularx}
\end{table*}

\subsection{Overall Performance Comparison}

We compare the performance of our proposed TPG with baselines mentioned above. Table~\ref{tab:2} reports the performance of TPG and eight baselines in terms of Recall@k and NDCG@k on five real world datasets. The ``Improv.'' column refers to the improvement rate of TPG compared to the second best model.
Based on the results, we observe that:

(1) Our proposed TPG significantly outperforms all the baseline methods on all datasets w.r.t. both NDCG@k and Recall@k and all values of k. Our proposed method achieves up to 20.2\% and 16.2\% improvements over the best-performing baseline in terms of NDCG@5 and Recall@5. It demonstrates the effectiveness and superiority of our proposed TPG, which makes effective usage of geographic information and temporal signal of the next location.

(2) Compared with RNN-based approaches, pure attention-based methods such as MobTCast, STAN, GeoSAN, and our proposed TPG clearly achieve better performances. It is reasonable since attention mechanism can capture global contextual information in spatial-temporal check-in sequences, while RNN-based methods suffer from the risk of forgetting past long-range information. Among RNN-based models, DeepMove and CARA generally have relatively better performances than others, which attributes to their consideration of spatial-temporal modelling, and short-term and long-term periodicity modeling. These designations make up for the inherent defects of RNN to a certain extent. Compared with attention-based state-of-the-art model MobTCast, STAN, and GeoSAN, the substantial improvement achieved by TPG demonstrates the importance of explicitly using temporal signal of the next location and shifted window mechanism for geo-gridding. Although STAN generally performs better than MobTCast and GeoSAN, it costs extremely large memory overhead and calculation time overhead due to matrix operations for all historical check-ins. Our method TPG is far superior to other methods in terms of computing time, memory, and accuracy.

(3) The density of check-in records for different datasets can be represented by \#check-ins $/$ (\#users $\times$ \#locations). The sparsities are 0.001, 0.007, 0.027, 0.009, and 0.120 for Gowalla, Brightkite, NYC, TKY, and SIN, respectively. This can explain why the improvement brought by TPG is larger in NYC and SIN than in TKY. Besides, it is obvious that TPG has a strong ability to handle sparse data like Gowalla.

\subsection{Interval Prediction Performances}

By introducing temporal prompts, TPG is able to make interval predictions with accurate timestamps of locations to be predicted. Relevant results are given in Table~\ref{tab:3}. We here mask one (``int. 1'' in Table~\ref{tab:3}), two (``int. 2'' in Table~\ref{tab:3}), and three (``int. 3'' in Table~\ref{tab:3}) most recent check-in(s) of users to test TPG and two baselines STAN and GeoSAN 's performances on all datasets. The detailed setting is using first 96, 97, 98 check-ins to predict the 100-th check-in in each user's trajectory.

For STAN and GeoSAN, we observe that compared with using all check-in data, the more the latest check-in(s) is/are masked, the more the performances drop marginally. It is reasonable since some previous studies \cite{pi2020search} have demonstrated that a user's recent behavior has a great impact on the user's next behavior. However, the situation is totally different for TPG's interval prediction. The performances are sometimes even better than the next location recommendation. These impressive results are very strong arguments for the benefit of explicit temporal information modeling by using timestamps as prompts.

\begin{table*}
\setlength{\tabcolsep}{0pt}
\centering
\caption{Performances of ablation studies.}
\label{tab:4}
\begin{tabularx}{\linewidth}{p{1.5cm}<{\raggedright}p{0.815cm}<{\centering}p{0.815cm}<{\centering}p{0.815cm}<{\centering}p{0.815cm}<{\centering}p{0.815cm}<{\centering}p{0.815cm}<{\centering}p{0.815cm}<{\centering}p{0.815cm}<{\centering}p{0.815cm}<{\centering}p{0.815cm}<{\centering}p{0.815cm}<{\centering}p{0.815cm}<{\centering}p{0.815cm}<{\centering}p{0.815cm}<{\centering}p{0.815cm}<{\centering}p{0.815cm}<{\centering}p{0.815cm}<{\centering}p{0.815cm}<{\centering}p{0.815cm}<{\centering}p{0.815cm}<{\centering}}
\toprule
& \multicolumn{4}{c}{Gowalla} & \multicolumn{4}{c}{Brightkite} & \multicolumn{4}{c}{NYC} & \multicolumn{4}{c}{TKY} & \multicolumn{4}{c}{SIN} \\ 
\cmidrule(r){2-5} \cmidrule(r){6-9} \cmidrule(r){10-13} \cmidrule(r){14-17} \cmidrule(r){18-21}
\multicolumn{1}{c}{} & R@5 & N@5 & R@10 & N@10 & R@5 & N@5 & R@10 & N@10 & R@5 & N@5 & R@10 & N@10 & R@5 & N@5 & R@10 & N@10 & R@5 & N@5 & R@10 & N@10 \\
\midrule
TPG & 63.2\% & 45.5\% & 74.6\% & 50.1\% & 62.4\% & 47.9\% & 74.4\% & 51.8\% & 37.3\% & 26.8\% & 52.8\% & 31.8\% & 65.4\% & 53.1\% & 76.5\% & 56.7\% & 19.4\% & 14.3\% & 34.3\% & 19.1\% \\
\midrule
\uppercase\expandafter{\romannumeral1}. $-$ TP & 60.4\% & 44.0\% & 73.0\% & 48.4\% & 61.1\% & 47.2\% & 73.6\% & 50.4\% & 34.3\% & 23.7\% & 47.1\% & 28.2\% & 62.2\% & 50.5\% & 72.9\% & 54.0\% & 18.3\% & 12.2\% & 29.7\% & 15.9\% \\
\uppercase\expandafter{\romannumeral2}. $-$ TE & 61.7\% & 44.0\% & 73.7\% & 49.0\% & 57.3\% & 43.7\% & 69.0\% & 47.5\% & 36.9\% & 25.0\% & 50.8\% & 29.4\% & 65.4\% & 53.3\% & 76.4\% & 56.9\% & 21.3\% & 15.9\% & 32.2\% & 18.4\% \\
\uppercase\expandafter{\romannumeral3}. $-$ SW & 58.77\% & 43.6\% & 72.6\% & 48.0\% & 61.5\% & 47.4\% & 73.7\% & 51.3\% & 36.8\% & 25.6\% & 52.3\% & 31.7\% & 64.0\% & 51.3\% & 75.1\% & 55.0\% & 19.9\% & 13.3\% & 30.5\% & 16.7\% \\
\uppercase\expandafter{\romannumeral4}. $-$ GE & 60.3\% & 42.6\% & 72.1\% & 47.6\% & 54.7\% & 42.2\% & 65.1\% & 45.6\% & 34.5\% & 25.2\% & 50.3\% & 28.50\% & 56.2\% & 43.3\% & 68.1\% & 47.4\% & 19.7\% & 14.3\% & 33.7\% & 17.1\% \\
\uppercase\expandafter{\romannumeral5}. $+$ UE & 62.1\% & 45.1\% & 74.0\% & 48.5\% & 57.9\% & 44.1\% & 70.1\% & 48.0\% & 36.3\% & 25.4\% & 51.1\% & 32.0\% & 53.2\% & 40.1\% & 65.9\% & 44.3\% & 18.5\% & 13.1\% & 26.7\% & 15.7\% \\
\bottomrule
\end{tabularx}
\end{table*}

\subsection{Ablation Study}

We conduct an extensive ablation study on TPG, to dive into the effectiveness of each module in our proposed model. It should be emphasized that our base model (denoted as TPG in the first row of Table~\ref{tab:4}) does not have user embedding. We consider the following variants of our model for ablation: 

\begin{itemize}
\item Remove TP (Temporal-based Prompt): We use locations as query of the decoder, instead of using temporal-based prompts.
\item Remove TE (Time Embedding): We remove time embedding of check-in sequences, only using POI embedding and geography embedding as inputs of history encoder.
\item Remove SW (Shifted Window Mechanism): We remove the shifted window mechanism in the geography-aware encoder.
\item Remove GE (Geography Encoder): We remove the geography encoder and only use POI embedding and time embedding as inputs of history encoder.
\item Add UE (User Embedding): We add user embedding into the history encoder by concatenating it with POI embedding, geography embedding, and time embedding.
\end{itemize}

The performance comparisons are shown in Table~\ref{tab:4}. From the comparisons, we have several findings:

(1) The overall performance of the model drops without temporal-based prompts, especially on Foursquare datasets with local regions. This phenomenon indicates the significance of explicitly incorporating temporal signal of the next location. Furthermore, results of ``Remove TE'' is generally slightly better than ``Remove TP'' in Gowalla Foursquare datasets, while still worse than TPG. It indicates that if time information is not used appropriately, it brings additional noise to the model and degrades the performances. It also proves that the temporal-based prompt strategy proposed in this paper makes effective use of time information.

(2) The results of ``Remove SW'' is worse than original TPG in most cases. It demonstrates that our proposed shifted window mechanism is an efficient augmentation method for spatial information. It can bridge the semantic gap of adjacent grids in terms of spatial distribution, and improves the accuracy and stability of the model. Performances of ``Remove GE'' is even worse than ``Remove SW''. It is reasonable since geographic information is of great significance in location-based applications.

(3) ``Add UE'' generally leads to great performance reduction compared with TPG. This is because adding user embedding to inputs of history encoder may contribute to the misaligned between the vector space of check-in sequences and the vector space of locations. This suggests future work to appropriately use user information to enhance the performance.

\begin{figure}
\centering
\includegraphics[width=\linewidth]{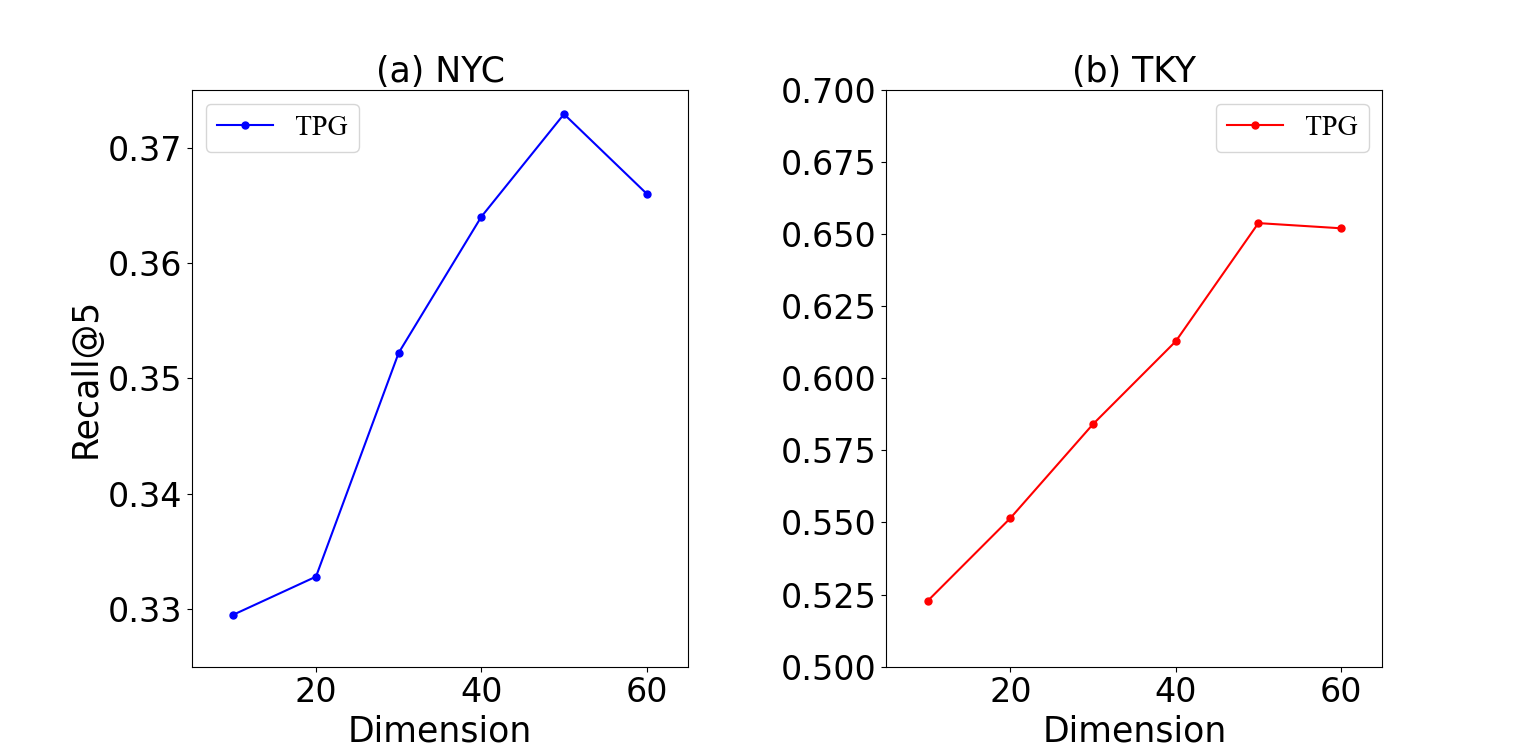}
\caption{The impact of geography embedding dimension for model performances.}
\label{fig:6}
\end{figure}

\subsection{Parameter Sensitivity Analysis}

We conduct sensitivity analysis on two important parameters, namely, geography embedding dimension and step size of the shifted windows. We first investigate model sensitivity with regard to geography embedding dimension. We vary the dimension used in the geography-aware encoder from 10 to 60 with a step of 10. The experimental results on two datasets NYC and TKY are reported in Figure~\ref{fig:6}. We can come to a conclusion that a small dimension for geography embedding will make the performance very poor. This is because small dimensions are difficult to describe the complex geographic relationships between POIs, which will cause great information loss. The model performance reaches the peak when the geography dimension is 50. When the dimension increases to 60, the performance decreases a bit. This may be explained that the size of the semantic space formed by the geography-aware encoder is certain. When the embedding dimension is too high, the information is unsaturated, and noise may be introduced instead.

\begin{figure}
\centering
\includegraphics[width=\linewidth]{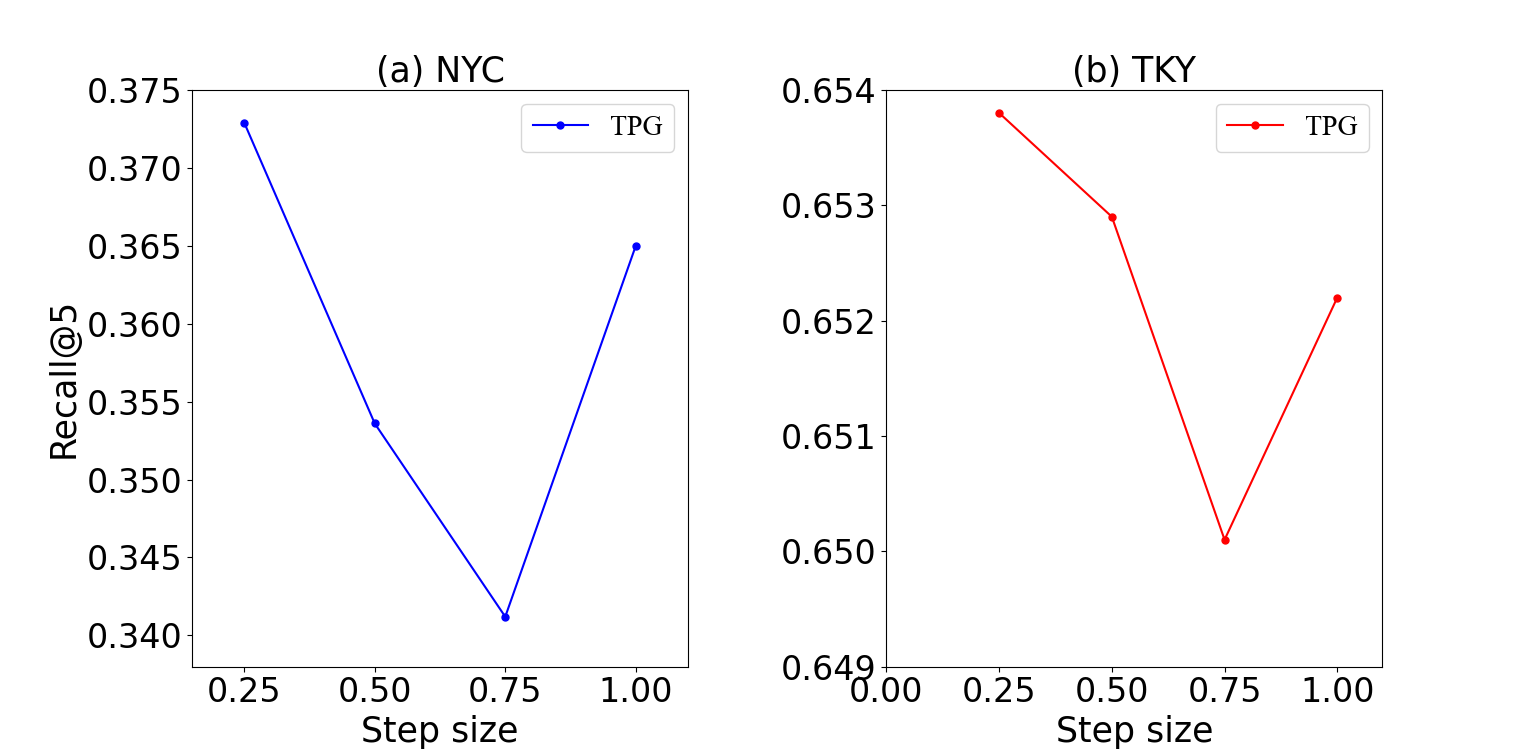}
\caption{The impact of the step size of shifted windows for model performances.}
\label{fig:7}
\end{figure}

We further investigate model sensitivity with regard to the step size of the shifted windows. We vary the step size used in the shifted window mechanism from 0.25 to 1 with a step of 0.25. Note that the step size here means the proportion of moving length and grid size. We still take two datasets NYC and TKY as examples. The experimental results are showed in Figure~\ref{fig:7}. We can find that the performance peaks at a small step size 0.25 for shifted window mechanism, and dropped until the step size is 0.75. Such phenomenon conforms to the First Law of Geography, which indicates ``everything is related to everything else, but near things are more related than distant things.'' We can also observe that when the step size is 1, there is a performance improvement. When the step size is 1, the model actually degenerates to directly aggregate the neighbor grids of the grid itself at each level. This phenomenon proves the defect of previous methods, that is, they do not fully consider the correlations of adjacent grids.

\section{Conclusion}

In this paper, we revisit the location recommendation problem. We find that most methods either ignore the prerequisite of knowing the exact time at which the POI needs to be predicted in real world applications, or implicitly fuse temporal information with historical check-ins. We propose TPG, a temporal prompt-based and geography-aware framework, for next location recommendations. We show how to use timestamps as prompts to explicitly model time information of locations to be predicted. By proposing a shifted window mechanism, we also show how to avoid the hard boundary problem with regard to geographic coordinates of check-ins. The experimental results on five benchmark datasets demonstrated the superiority of TPG compared with other state-of-the-art methods. The results indicated that temporal signals of locations are of great significance. We also demonstrate through ablation studies that our proposed shifted window mechanism is capable of overcoming defects with regard to geographic information modeling of previous approaches.

As for future work, we plan to design more intelligent prompts. Large-scale pre-trained models from NLP communities have demonstrated unlimited potential of prompt learning. We can consider combining location recommendation with language pre-trained models through prompts.

\bibliographystyle{ACM-Reference-Format}
\bibliography{main}

\end{document}